\documentclass[a4paper,11pt]{aretthesis}

\usepackage[estonian, english]{babel}
\usepackage[utf8]{inputenc}

\usepackage[T1]{fontenc} 
\usepackage{times}

\usepackage{longtable} 
\usepackage{aas_macros} 
\usepackage{verbatim} 
\usepackage{amsmath} 
\usepackage{amsfonts} 
\usepackage{amssymb}  
\usepackage{amscd} 
\usepackage{graphicx,epsfig} 
\usepackage{subfig} 
\usepackage{float} 
\usepackage{lscape} 
\usepackage{url}
\usepackage{sidecap}
\sidecaptionvpos{figure}{t}

\usepackage[english,estonian]{babel}

\usepackage{accents}

\usepackage{natbib}

\newcommand{\bn}{\begin{enumerate}}
\newcommand{\en}{\end{enumerate}}

\newcommand{\bed}{\begin{displaymath}}
\newcommand{\eed}{\end{displaymath}}

\newcommand{\ie}{{\it i.e. }}

\def\Mp{\,{\rm M_{\odot} /pc^2}}

\def\Mp3{\,{\rm M_{\odot} /pc^3}}

\newcommand{\be}{\begin{equation}}
\newcommand{\ee}{\end{equation}}
\newcommand{\dd}[1]{{\rm d}#1\,}

\newcommand{\ba}{\begin{array}} 
\newcommand{\ea}{\end{array}}

\def\araa{ARA\&A}%
\def\apj{ApJ}%
\def\apjl{ApJ}%
\def\apjs{ApJS}%
\def\ao{Appl.~Opt.}%
\def\aap{A\&A}%
\def\jcap{J. Cosmology Astropart. Phys.}%
\def\mnras{MNRAS}%
\def\bain{Bull.~Astron.~Inst.~Netherlands}%
\bibpunct{(}{)}{;}{a}{}{,} 

\begin{document}

{\large
  {\bf The question of selective absorption of light in space viewed
  from the viewpoint of the dynamics of the
  universe\footnote{\footnotetext ~~Published by \citet{Opik:1915}}
  }
   
\medskip
 
{\large
{\em {E. \"Opik}}
}

\medskip

\begin{quote}
%  {\small
{\bf Abstract.}
The selective light absorption in space has been raised in
astronomical literature. The substance producing the absorption must
have some mass; thus the question is how large it is. We develop a
dynamical model of the Milky Way system, assuming that it can be
represented by a flattened ellipsoid of rotation. We use the spatial
distribution of $\delta$-Cephei and Algol type variable stars, and
mean velocities of stars according to Campbell to calculate the
dynamical density of the Milky Way near the Sun,
$0.100\,\Mp3$.  We find that the 
dynamical density is equal to the mean density of stars in the
vicinity of the Sun.  Our conclusion is that the
intrinsic gravity of stars fully explains their motion, and the
existence of any other matter in any significant quantity seems
unlikely. Therefore, the existence of noticeable selective absorption
seems to be absolutely improbable, unless one admits the existence in
the space of particles much smaller than atoms of elements known to
us.  Normal absorption may exist if the particle diameter is of the
order of a millimetre or less, and their mass is comparatively
small. This absorption has not yet been reliably detected; the fact
that the number of stars increases with stellar magnitude more slowly
than theory requires in case of uniform distribution of stars in
space, can be equally explained by both light absorption and decrease
in number of stars with  distance.

\footnote{Abstract and Keywords added by  editors 
  of the English translation Jaan Einasto and Peeter Tenjes.}.
%}
\end{quote}

\medskip

Recently, in astronomical literature, the question of light absorption
in space has been raised with increasing frequency, mostly selective
absorption, \ie absorption that is stronger the smaller the
wavelength; a review of work on determining the magnitude of selective
absorption was recently given by \citet{Kapteyn:1914tx}.  The
existence of absorption has not yet been proved with certainty, and
how shaky may be the grounds for its determination, is shown by the
work of \citet{Adams:1914vy}, which demonstrates that the well-known
fact of a decrease of the violet end of the spectrum with the decrease
of the proper motion, \ie with the increase of the star's distance, is
not confirmed yet. Namely, with equal apparent brightness the more
distant stars are absolutely brighter and bigger in their sizes and,
in all probability, are surrounded by accordingly thicker atmosphere.

I would like to draw attention to one side of the question, which, as
far as I know, has not been covered so far. The substance producing
the absorption must have some mass; the question is how large it
is. As it is known, selective absorption is created by particles small
in size in relation to the wavelength, such as gas molecules; the
amount of the absorption depends only on the number of these molecules
per unit volume. Let $\mu$ be the average molecular weight (relative
to air) of the substance filling the space, and $a$ -- the visual
absorption coefficient expressed in magnitudes per light-year, $\rho$
-- the density of the absorbing matter (relative to the Sun).  Based
on \citet{Abbot:1911wb} measurements at Mount Whitney at pressure
$p = 440$ mm the coefficient of transparency for visual rays
($\lambda= 550\, \mathrm{nm}$) is equal to 0.918, which corresponds to
the absorption of 0.070 magnitudes.  The air mass at Mt. Whitney is
equal to 6000 kg per square meter; in such a case the mass of
interstellar matter inside a parallelepiped, whose base is 1 square
metre and whose length is 1 light year, is $6000\, a\mu/0.07$~kg;
because 1 light year $= 9.45 \times 10^{15}$ meters and the density of
the Sun is 1410 kg/m${^3}$, it is not difficult to calculate that the
density of the interstellar substance is equal to
\begin{equation}
\rho = 6.4\times 10^{-15} a\mu ~~ (\odot = 1) .	\label{eq1}
\end{equation}

Let $\delta$ be the ``density of stellar matter in space'', \ie the
density of some imaginary matter, which, being evenly distributed in
space, will have the mass equal to the mass of encapsulated in this space 
stars.  It is not difficult to determine the order of magnitudes of $\delta$.
According to Kaptein's studies, within a sphere of $r= $ 500
light  years there are about $1.5\times 10^6$ stars; let their average mass 
($\odot = 1$) be $\overline{m}$, then  (as the radius of the Sun is $0.72\times
10^{-7}$ light years)
\be
\delta = \frac{1.5\times 10^6 (0.72\times 10^{-7})^3}{500^3}
\overline{m} = 0.44\times 10^{-23} \overline{m} . 	 \label{eq2}
\ee

Comparing $\rho$ and $\delta$, we see that even at the lowest value of
$a$ the former significantly exceeds the latter.   Indeed, taking $a=
0.0001$ (most investigators give larger values), $\mu=1/14$ 
(molecular weight of hydrogen), we  obtain
$$\rho = 4.7\times 10^{-18} ;$$
the quantity $\overline{m}$ is close to unity, and
$$\delta = 0,44 \times 10^{-23}. $$
Thus, at absorption of 1 magnitude per 10,000 light-years the density,
and consequently, also the mass of the absorbing substance, must be at least
100\,000 times greater than the mass of all stars! Obviously, in such
a distribution of absorbing substance in  space the movements of the stars
should depend almost entirely not on their mutual gravity, but on the
attraction of this absorbing matter scattered in space. However, such a large
value of the mass of the interstellar matter seems very unlikely --  the velocities of
stars would accordingly be much larger than the observed values, and
would be measured in thousands, not tens, of kilometres per second.  Therefore we
have to choose between the following suppositions (or to admit them
both): 1) the value of selective absorption is much lower than is
commonly assumed; 2) the molecular weight of the substance causing
selective absorption is much lower than the molecular weight of
hydrogen.  This strongly suggests the idea of electrons scattered in
space.

Since even the most negligible selective absorption must be dominated
by interstellar matter,  it is of some interest to find answer to the
question about the laws of motions  inside a 
star system, enveloped by  evenly dispersed nebular matter of 
considerable mass. (The question on dynamics of a
whole stellar system has already been developed by \citet{Eddington:1913wz}
for a spherical system, consisting of only
stars). Our main task will be to determine the order of matter
density, 
filling the Milky Way system.   Although, as it will turn out
later, our assumptions will not correspond to the truth precisely,
however, they will not influence  the order of magnitude of the
sought density.

Let us make the following assumptions: the interstellar matter is distributed as
%collected in a 
an ellipsoid of rotation of indefinite dimensions,
%speed,
uniform density $\rho$, and with the axial ratio $q$.  Inside the ellipsoid there are
stars scattered within it, the stellar density $\delta$ (it is
proportional to the number of stars per unit volume)  reaches its maximum
value at
the ellipsoid centre, and  the surfaces of equal density $\delta$  (the
surface level) are ellipsoids,  
similar and concentric to the above mentioned one.  If the equation of such
a surface is
$$q^2x^2 + y^2 + z^2 = u^2 $$
(the minor axis of the ellipsoid lies on the $x$-axis), then the stellar
density is
\be
\delta  = f (u^2). \label{eq3a} 
\ee
At last, we assume that $\delta$ is significantly smaller than $\rho$, so the mass
of the stars can be neglected.

The gravitational acceleration at a given point inside the ellipsoid of
rotation of uniform density $\rho$ is expressed by the following
formulae:
\be
\left.
\begin{array}{ll}
\frac{\dd{^2x}}{\dd{t^2}}  = & -cx \\
\noalign{\medskip}
\frac{\dd{^2 y}}{\dd{t^2}}  = &-c_1y \\
\noalign{\medskip}
\frac{\dd{^2 z}}{\dd{t^2}}  = &-c_1z 
\end{array}
\right\} ,
\label{eqA}
\ee
where
\be
\left.
\ba{ll}
c  = &4\pi n \rho \frac{q^2}{(q^2 -1)^{3/2}} \left( \sqrt{q^2 -1} -\arctan \sqrt{q^2-1}\right) \\
\noalign{\medskip}
c_1 = & 2\pi n\rho \frac{q^2}{(q^2 -1)^{3/2}} \left( \arctan
  \sqrt{q^2-1} - \frac{\sqrt{q^2-1}}{q^2} \right) ; 
\ea
\right\}
\label{eqB}
\ee
here $q$ is the axis ratio of the ellipsoid, and $n$ is the gravitational constant.

It is clear from this that the motion projections on coordinate axes
are harmonic oscillations, and the oscillation term for $x$-axis is
different than for $y$-axis and $z$-axis. Having solved these equations,
they can be reduced to the following form:
\be
\left.
\ba{ll}
x  = &H \cos (t\sqrt{c} ) \\
\noalign{\smallskip}
y  = &K \cos (t\sqrt{c_1} -\eta ) \\
\noalign{\smallskip}
z  = &L \cos (t\sqrt{c_1} -\zeta ) 
\ea
\right\} .
\label{eq3}
\ee
$H$, $K$ and $L$ are obviously the values of the greatest distance of a given
point mass along the coordinate axes, $\eta$ and $\zeta$ are the phase
differences of the oscillations. Let us call $H$, $K$, and $L$ the amplitudes
along the coordinate axes.

Let us denote by $u$, $v$ and $w$ the velocity
components along the coordinate axes; then it is easy to obtain the
following relations from equations (\ref{eq3}):
\be
\left.
\ba{ll}
u  = &\pm \sqrt{c} \sqrt{H^2 - x^2} \\
\noalign{\smallskip}
v  = &\pm \sqrt{c_1} \sqrt{K^2 - y^2} \\
\noalign{\smallskip}
w  = &\pm \sqrt{c_1} \sqrt{L^2 - z^2} 
\ea
\right\} .
\label{eq4}
\ee

Let the number of stars whose amplitudes along the $x$-axis are enclosed between
$H$ and $H + \dd{H}$ be
\be
\dd{n_H} = \varphi (H) \dd{H}, 
\label{eq5}
\ee
and similarly for other axes
\be
\dd{n_K} = \psi (K) \dd{K}, 
\label{eq5a}
\ee
\be
\dd{n_L} = \psi (L) \dd{L}. 
\label{eq5b}
\ee

The functions $\varphi$ and $\psi$ give the  distribution functions of amplitudes along the
axes; for the $y$ and $z$-axis, identical  function is assumed due to
symmetry. Let us determine what form these functions should take in
order for the system to be in equilibrium at a given $\delta=f(u^2)$. For
this purpose we define the  number of stars whose abscissaes are between
$x$ and $x+\dd{x}$, $y+\dd{y}$ and $z+\dd{z}$.  On the one hand  this number is
obviously
\be
\left.
\ba{ll}
\dd{n_x}  = &\dd{x} \iint \delta\, \dd{y} \dd{z} \\
\noalign{\medskip}
\dd{n_y}  = &\dd{y} \iint \delta\, \dd{x} \dd{z} \\
\noalign{\medskip}
\dd{n_z}  = &\dd{z} \iint \delta\, \dd{x} \dd{y} 
\ea
\right\} .
\label{eq6}
\ee
The integration limits for an unbounded stellar system are constant;
the lower limits are 0, the upper limits -- infinity.

On the other hand, $\dd{n_x}$ depends on the form of the function
$\psi(H)$.   We note
that for the same star its $x$ coordinate varies in time from 0 to $H$ 
(pay attention only to the absolute value of $x$). 
Therefore at a given distance $x$ there can be only these stars for which
$H \ge x$; the time at which a given star changes  $x$  from $x$ to $x+ \dd{x}$ will
be equal to
$$\dd{t} = \frac{\dd{x}}{|u|}. $$
Consider the number of stars with the same $H$; obviously, the fraction of
these stars between $x$ and $x + \dd{x}$, will constitute on average as much
of the total number as the fraction of the total time of  the
oscillation  has $\dd{t}$, \ie  this fraction is equal to
$$ \frac{\dd{t}}{\tau_x} = \frac{\dd{x}}{|u| \tau_x} ,$$
where $\tau_x=2\pi/\sqrt{c}$ is the period of oscillation, and $|u|=\sqrt{c}\sqrt{H^2-x^2}$.
Substituting these variables we obtain that the number of
stars with a given $H$, enclosed between $x$ and $x + \dd{x}$,  is $\dd{x}/\sqrt{H^2-x^2}$;
but the number of stars with the given $H$ is $\varphi(H)\dd{H}$, 
therefore we obtain
\be
\dd{n_x} = \dd{x} \int_{H=x}^{H=\infty} \frac{\varphi(H)\, \dd{H}}{\sqrt{H^2-x^2}}. 
\label{eq7}
\ee

Comparing formulae (\ref{eq6}) and (\ref{eq7}) we obtain
\be
\iint \delta\, \dd{y} \dd{z} = \int \frac{\varphi(H)\, \dd{H}}{\sqrt{H^2-x^2}}. 
\label{eq8}
\ee

This equation can be called an equation of state of a given system;
for a given $\delta$ we can determine $\varphi$  and vice versa. 

Let the stellar density law be similar to the normal distribution
%law of chance,
\ie  let  
\be
\delta_0 = \delta e^{-k^2 (q^2x^2 + y^2 + z^2)} .
\label{eq9}
\ee

This law seems rather probable for the Milky Way stellar system;
moreover, we do not care about the exact form of the function, but it
is only important that it satisfies some general conditions -- that it
decreases with distance, the level surfaces are ellipsoids, and the
system is unbounded.

Substituting (\ref{eq9}) into (\ref{eq8}) and integrating, we obtain
\be
\int_x^\infty \frac{\varphi(H)\, \dd{H}}{\sqrt{H^2-x^2}} = \frac{A}{k^2} e^{-k^2q^2x^2} ,
\label{eq10}
\ee
where
$$A=\frac{\pi\delta_0}{4} .$$

The first part of this equation can be represented in the form:
$$
\int_x^\infty \frac{\varphi(H) \dd{H}}{\sqrt{H^2-x^2}} = \left[ \frac{\varphi(H) \sqrt{H^2-x^2}}{H} \right]_{H=x}^{H=\infty} - $$
$$ 
- \int_0^\infty \sqrt{H^2-x^2} \frac{[H\varphi '(H) -\varphi(H)] \dd{H}}{H^2} ;
$$
suppose that when
$$H=\infty ~~~ \varphi(H)=0;$$
then, as it is easy to see, the first term in the right side of
the equality will be 0, and we will have (taking into account equality
(\ref{eq10})) 
$$
-\int_x^\infty \sqrt{H^2-x^2} \left[  \frac{[H\varphi '(H) -\varphi(H)]}{H^2} \right] \dd{H} = \frac{A}{k^2} e^{-k^2 q^2 x^2} . $$
Differentiate this equation by $x$, applying the formula
$$
\left( \frac{\dd{}}{\dd{a}} \int_a^b f(x,a) \dd{a} = -f(a,a) + \int_a^b f'_a (x,a) \dd{a} \right) ,
$$
then we will have after a reduction by $x$:
$$
\int \frac{[H\varphi '(H) -\varphi(H)]\dd{H}}{H^2 \sqrt{H^2-x^2}} = -2Aq^2 e^{-k^2q^2x^2} , 
$$
or by excluding $e^{-k^2q^2x^2}$ with formula (\ref{eq10}) and transferring everything to 
the left-hand side, we will have
$$
\int_x^\infty \frac{1}{\sqrt{H^2-x^2}} \left[ 2k^2q^2\varphi(H) - \frac{\varphi(H)}{H^2} + \frac{\varphi '(H)}{H} \right] \dd{H} = 0.
$$
This identity is correct when
$$
\left( 2k^2q^2 -\frac{1}{H^2} \right) \varphi(H) + \frac{\varphi '(H)}{H} = 0,
$$
or
$$
\frac{\varphi '(H)}{\varphi(H)} = -2k^2q^2 H + \frac{1}{H} ;
$$
hence by a simple integration we find
\be
\left.
\ba{ll}
\varphi(H)  = &CH e^{-k^2q^2H^2} \\
\noalign{\medskip}
\psi(K)  = &C\frac{K}{q} e^{-k^2K^2} \\
\noalign{\medskip}
\psi(L)  = &C \frac{L}{q} e^{-k^2L^2 }
\ea
\right\} .
\label{eq11}
\ee

These formulae give the  amplitude distribution function.  It is
not difficult to determine the average value of the velocity
components of the stars along the axes for any point. Let's find the
average absolute value of the velocity component parallel to the
$x$-axis for a certain $x$-value; it will obviously be equal to
$$
\overline{u}_x = \frac{ \int_x^\infty
  \frac{u\varphi(H)\dd{H}}{\sqrt{H^2-x^2}} }{  \int_x^\infty
  \frac{\varphi(H)\dd{H}}{\sqrt{H^2-x^2}} } ;
$$
substituting
$$ u = \sqrt{c} \sqrt{H^2-x^2} ,$$
we find after the integration
\be
\overline{u}_x = \frac{\sqrt{c}}{kq\sqrt{\pi}} .
\label{eq12}
\ee

We see that the average value of the velocity component of stars under
the chosen density function  is a constant for all points of the system. For
the average value of the velocity components on the other 
axes we will get the same expression:
\be
\overline{u}_y = \overline{u}_z = \frac{\sqrt{c_1}}{k\sqrt{\pi}}. \label{eq12a}
\ee

The average absolute value of the velocity can be assumed to be
(approximately)
\be
V = \sqrt{ \overline{u}^2_x + \overline{u}^2_y + \overline{u}^2_z} = \frac{1}{k\sqrt{\pi}} \sqrt{\frac{c}{q^2} + 2c_1} .
\label{eq13}
\ee

The constants $c$ and $c_1$, as seen from formulae (\ref{eqB}), depend on the
density and shape of the system; $k$ determines the rate at which the
number of stars decreases with distance from the centre. For a
spherical body, at $q= 1$,  the following expression is obtained for
$c$ instead of (\ref{eqB}):
$$
c = \frac{4}{3} \pi n\rho .
$$
Let us choose the constant $n$ so that, if the unit of
length is kilometre and the unit of time is second, then the unit of 
the density would be the solar density;  we have
$$
\frac{\dd{^2x}}{\dd{t^2}} = - \frac{4}{3} \pi n\rho x,
$$
and for the Sun
$$\rho = 1, ~~~ x= 695000, ~~~ \frac{\dd{^2x}}{\dd{t^2}} = 0.274; $$
from here
$$ 4\pi n = 1.2\times 10^{-6} .$$

Let us now try to apply the results of our reasoning to the Milky Way.
According to the currently known data, it is a flat disc or a
flattened ellipsoid of rotation with a ratio of axes approximately
equal to 10. For determination of the approximate value of the constant $k$
we use the results by \citet{Hertzsprung:1913vo} 
and \citet{Russell:1914tj}, 
obtained from the examination of the spatial distribution of variable
stars of $\delta$-Cephei and Algol type.  It turns out, that these
stars are 
heavily concentrated to the plane of the Milky Way, with average
distance from the plane for type $\delta$-Cephei at 260 light years, for Algol at
440 light years; the difference between these two numbers could be real, but
could be also due to some systematic reasons, depending on some
accepted assumptions. It is possible to accept an average value
of $\overline{x}= 350$ light years or $3.3 \times 10^{15}$ km. On the other hand, it follows
from formula (\ref{eq9}) that with fixed $y$ and $z$ values the stellar density
(number of stars) is expressed by the formula
$$ 
\delta = C e^{-k^2q^2x^2} ; 
$$
their average distance from the plane of the Milky Way will be
$$
\overline{x} = \frac{ \int_x^\infty x e^{-k^2q^2x^2} \dd{x} }{  \int_x^\infty e^{-k^2q^2x^2} \dd{x}  }  = \frac{1}{kq\sqrt{\pi}} .
$$
Accepting
$$ q=10, ~~~ \overline{x} = 3.3\times 10^{15}, $$
we found
$$ k = 1.72\times 10^{-17} . $$

Then at $q = 10$, $4\pi\,n = 1.2 \times 10^{-6}$ our constants $c$ and
$c_1$, will be $10^{-6}\rho$  and $0.08 \times 10^{-6}\rho$, respectively.

Substituting these values into formula (\ref{eq13}), and taking according to 
\citet{Campbell:1915tz} $V=30$~km/s, we find
$$\rho=0.48\times 10^{-23}$$
in  units of solar density\footnote{ Since solar density is 1410 kg/m$^3$ and 
$1 \Mp3 = 6.770 \times 10^{-20} \, \mathrm{kg/m^3}$,, we get 
the density  in solar masses per cubic parsec:
$\rho = 0.100\,\Mp3$ (Note added by  editors 
  of the English translation Jaan Einasto and Peeter Tenjes).}.

Consequently, the effective density for the hypothetical homogeneous
medium turns out to be exactly equal to the ``stellar density'', which
was calculated in the beginning of this paper.  From this it is clear
that the intrinsic gravity of stars fully explains their motion and
the existence of any other matter in any significant quantity seems
unlikely.  Therefore, the existence of noticeable selective absorption
seems to be absolutely improbable, unless one admits the existence in
the space of particles much smaller than atoms of elements known to
us.

In addition to selective absorption, normal absorption equal to all
wavelengths can exist, produced by particles with a diameter larger than
the wavelengths. If these particles are completely opaque,
$d$ their thickness in millimetres, then, taking their density equal to
the density of water, it is easy to obtain the following expression
for the effective density of the space filled with them:
$$\rho=0.75~ a d\times 10^{-19} , $$
where $a$ is, as before, the absorption coefficient in magnitudes per
1 light year. At $a = 0.0001$ we have
$$\rho=0.75~ d\times 10^{-23} .$$

Consequently, normal absorption may well exist if the particle
diameter is of the order of a millimetre or less, and their mass is
comparatively small. However, it must be remembered that this
absorption has not yet been reliably detected; the fact that the
number of stars increases with stellar magnitude more slowly than
theory requires in case of uniform distribution of stars in space, can
be equally explained by both light absorption and decrease in number
of stars with  distance.

\medskip
Moscow University

30 April, 1915
\medskip

\setcounter{equation}{0}
{\bf Comment added by editors Jaan Einasto and
  Peeter Tenjes}. \\
\\
The paper by \citet{Opik:1915} is  the first one where the problem of
invisible (dark) matter in solar neighbourhood of Galaxy was
discussed. His results were confirmed by \citet{Kapteyn:1922wu}, who
introduced the term ``dark matter'' to denote the possible invisible
matter in the solar neighbourhood. This problem was 
studied by \citet{Jeans:1922}, \citet{Oort:1932tb} and 
in Tartu Observatory by 
\citet{Kuzmin:1952ab, Kuzmin:1955aa}, \citet{Eelsalu:1958aa} and
\citet{Joeveer:1972, Joeveer:1974wa, Joeveer:1975vs}.

The dynamical density $\rho$ can be
calculated from the gradients of the gravitational potential $\Phi$ via the
Poisson equation, which has in  cylindrical coordinates  the form
\be
4\pi\,G\rho=-\frac{\partial^2\Phi}{\partial\,z^2}-\frac{\partial^2\Phi}{\partial\,R^2}
- \frac{1}{R}\frac{\partial\,\Phi}{\partial\,R},
\label{eq1.5}
\ee
where $G$ is the gravitational constant.  To determine the dynamical density
\citet{Kuzmin:1952ab} made several innovations compared with earlier
studies. First, he noticed 
that planar subsystems rotate practically with a circular velocity.
In this case we can express the sum of second and third terms in
Eq.~(\ref{eq1.5}) near the Galactic plane through Oort galactic rotation
parameters $A$ and $B$:
    \be
    \frac{\partial^2\Phi}{\partial R^2}
      +\frac{1}{R}\frac{\partial\Phi}{\partial
        R}=-2\frac{\dd{V_c}}{\dd{R}}\frac{V_c}{R}=2(A^2-B^2),
\label{eq1.51}
\ee
where $V_c$ is the circular velocity. 
Kuzmin introduced for the first term of Eq.~(\ref{eq1.5}) the designation
$$C^2 = -(\partial^2\Phi/\partial\,z^2)_{z=0}.$$
The parameter $C$ has the same dimension as $A$ and $B$, and is a
necessary complement to the rotational Oort parameters.
Assuming that  velocities $v_z$ of a stellar population have normal distribution 
Kuzmin showed that near the Galactic plane also their $z$ coordinates have normal 
distribution, and the parameter $C$ can be calculated from the ratio
of the dispersions of vertical velocities $\sigma_z$ and spatial
$z$-coordinates $\zeta$:
\be
C={\sigma_z \over \zeta}.
\label{eq1.48}
\ee
In other words, to derive the dynamical density, it is not needed to
calculate the gravitation attraction, $K_z=-\partial\Phi/\partial\,z$,
over a large $z$-interval,  as 
done by \citet{Kapteyn:1922wu}, \citet{Jeans:1922} and \citet{Oort:1932tb}, but
only its gradient near the Galactic plane. Since the gradient changes
near the Galactic plane rather quickly, only  flat stellar
populations are suitable to find dispersions $\sigma_z$ and
$\zeta$.

The second innovation  by \citet{Kuzmin:1952ab} was to use for
determination of dispersions of spatial position and velocities  of
{\em identical
samples of  A and gK stars near the galactic plane within galactic latitudes
$\pm 3$ degrees}.  To find the velocity  dispersion $\sigma_z$ he used
vertical components of proper motions, to calculate spatial dispersion
$\zeta$ he used parallaxes. This eliminates possible sampling errors and
errors in parallaxes. His final result \citep{Kuzmin:1955aa} was
\be
C = 68 \pm 3~~{\rm  km/sec/kpc}. \nonumber
\ee
and for the density:
\be
\rho= (5.2 \pm 0.5) \times 10^{-24} {\rm g/cm}^3 = 0.077 \pm 0.008\,
\Mp3. \nonumber
\ee
This value confirmed \citet{Opik:1915} and \citet{Kapteyn:1922wu}  conclusion that the gravity of
known stars fully explains their motion, and there is no need for
local dark matter.  Further determinations of $C$ were made in Tartu
by \citet{Eelsalu:1958aa}, and
\citet{Joeveer:1972, Joeveer:1974wa, Joeveer:1975vs}.
These studies by  Kuzmin, Eelsalu and J\~oeveer 
formed their PhD theses (candidate theses according to Soviet rules).
The final result of the Eelsalu analysis is: $C=67 \pm 3$~km/sec/kpc.
The mean value of the J\~oeveer analysis is $C=70$ ~km/sec/kpc, and
$\rho_{dyn}=0.09\, \Mp3$. 

Different results were  obtained by \citet{Oort:1960uy} and
  \citet{Bahcall:1984wl}, \citet{Bahcall:1992wb}. Their analyses  suggested the
presence of local dark matter approximately in the same quantity as
the known matter.  This discrepancy led to a discussion, for 
overviews see \citet{Gilmore:1989}, \citet{Read:2014wo} and
\citet{McKee:2015ug}.  \citet{Kuijken:1989td} made a careful analysis
of the \citet{Bahcall:1984wl} study and found flaws in it.
\citet{Kuijken:1989td} made a reanalysis of F dwarf and gK giant stars
and found no evidence for any missing matter near the Sun. 

Flat rotation curves of galaxies (\citet{Rubin:1978},
\citet{Bosma:1978}) suggest the presence in galaxies almost spherical
extended coronas of dark matter. Models of the Galaxy including
coronas were developed by \citet{Einasto:1976vu},
\citet{Bienayme:1987wt} and \citet{Haud:1989fo}. These models predict
the presence of local dark matter in solar vicinity with density of
the order $\rho_{DM}\simeq 0.01~\Mp3$.

Recent determination of the
density of matter in solar vicinity have led to values $0.097 \pm
0.013\, \Mp3$, the estimated sum of the stellar and gas
mass densities $\rho_\star=0.084\pm 0.012\, \Mp3$, which
yield for  the density of dark matter
$\rho_{DM}=0.013 \pm 0.003\, \Mp3$ (\citet{Bienayme:1987wt,
  Bienayme:2006uz}, \citet{Creze:1998uu},  \citet{Holmberg:2000vq},
\citet{McKee:2015ug}). Local matter density determinations based on the Gaia
satellite data fully confirm these results (\citet{Kipper:2018va}, \citet{Buch:2019}, 
\citet{Guo:2020}, \citet{Salomon:2020}).

%\bibliography{opik15}{} %opik15.bib
\bibliographystyle{aa2}

\end{document}